\documentclass[10pt]{iopart}
\usepackage{iopams} 
\usepackage{gensymb}
\usepackage{graphicx}

\usepackage{subcaption}
\usepackage{amsbsy}
\usepackage{color}
\usepackage{textcomp}
\usepackage{float}
\usepackage{sidecap}
\usepackage{siunitx}
\usepackage{amssymb}

\expandafter\let\csname equation*\endcsname\relax

\expandafter\let\csname endequation*\endcsname\relax
\usepackage{amsmath}
\usepackage{makecell}
\usepackage{upgreek}

\protected\def\PRE{\ifmmode \mathrm{PRE} \else PRE\fi}

\begin{document}

\title{I-mode pedestal relaxation events at ASDEX Upgrade}

\author{D. Silvagni${^{1,2}}$, T. Eich${^1}$, T. Happel${^1}$, G. F. Harrer${^3}$, M. Griener${^1}$,  M. Dunne${^1}$, M. Cavedon${^1}$, M. Faitsch${^1}$, L. Gil${^4}$, D. Nille${^1}$, B. Tal${^1}$, R. Fischer${^1}$, U. Stroth${^{1,2}}$, D. Brida${^1}$, P. David${^1}$, P. Manz${^{1,2}}$, E. Viezzer${^5}$, the ASDEX Upgrade team$^{\mathrm{a}}$ and the EUROfusion MST1 team$^{\mathrm{b}}$}

\address{$^1$Max-Planck-Institut f{\"u}r Plasmaphysik, Boltzmannstr. 2, 85748 Garching, Germany \\ $^2$Physik-Department E28, Technische Universit{\"a}t M{\"u}nchen, James-Franck-Str. 1, 85748 Garching, Germany \\ $^3$Institute of Applied Physics, TU Wien, Fusion@{\"O}AW, Wiedner Hauptstr. 8-10, 1040 Vienna, Austria\\ $^4$Instituto de Plasmas e Fus{\~a}o Nuclear, Instituto Superior T{\'e}cnico, Universidade de Lisboa, 1049-001 Lisboa, Portugal \\ $^5$Dept. of Atomic, Molecular and Nuclear Physics, University of Seville, Avda. Reina Mercedes, 41012 Seville\\ $^{\mathrm{a}}$see author list of H. Meyer et al. 2019 Nucl. Fusion 59 112014
 \\ $^{\mathrm{b}}$see author list of B. Labit et al. 2019 Nucl. Fusion 59  086020}
 
\ead{davide.silvagni@ipp.mpg.de}
\vspace{10pt}




\begin{abstract}
The I-mode confinement regime can feature small edge temperature drops that can lead to an increase in the energy deposited onto the divertor targets. In this work, we show that these events are associated with a relaxation of both electron temperature and density edge profiles, with the largest drop found at the pedestal top position. Stability analysis of edge profiles reveals that the operational points are far from the ideal peeling-ballooning boundary. Also, we show that these events appear close to the H-mode transition in the typical I-mode operational space in ASDEX Upgrade, and that no further enhancement of energy confinement is found when they occur. 
Moreover, scrape-off layer transport during these events is found to be very similar to type-I ELMs, with regard to timescales ($\approx 800\,\upmu$s), filament propagation, toroidally asymmetric energy effluxes at the midplane and asymmetry between inner and outer divertor deposited  energy. In particular, the latter reveals that more energy reaches the outer divertor target. Lastly, first measurements of the divertor peak energy fluence are reported, and projections to ARC -- a reactor designed to operate in I-mode -- are drawn. 
\end{abstract}

%
%
%
%
\ioptwocol

\section{Introduction}

The I-mode is an attractive operational regime for magnetically confined fusion reactors. It is characterized by the high-energy confinement of H-mode and the relatively poor particle confinement of L-mode ~\cite{Ryter_1998, Whyte_2010}. As a consequence, a temperature pedestal is formed at the plasma edge, while density edge profiles remain similar to those in L-mode. In this way, the I-mode simultaneously combines the desired properties of L-mode and H-mode plasmas, namely reduced impurity and helium ash accumulation in the plasma core, and large fusion energy production. Moreover, the I-mode is free of type-I edge localized modes (ELMs), as its pedestal is ideal peeling-ballooning stable~\cite{Walk_2014, Happel_2017}. It is generally recognized that a large number of type-I ELMs will not be acceptable for ITER and next-step fusion devices~\cite{Eich_2017}. Hence, for these reasons the I-mode is considered as a possible operational regime for a fusion power plant. 
\newline
However, I-mode can feature small edge temperature drops that can ultimately lead to a consistent increase in the energy deposited onto the divertor targets. These I-mode ``ELM-like'' events were first observed in Alcator C-Mod~\cite{Walk_2014}. Similar events were also seen in DIII-D~\cite{Hubbard_2016} and more recently in ASDEX Upgrade (AUG). Understanding the appearance and the amount of energy expelled by these events is fundamental for the I-mode candidature as an operational regime in a fusion power plant. 
\newline It is worth clarifying that these events exhibit macroscopic differences with respect to the previously studied I-mode bursts in AUG~\cite{Happel_2016, Happel_2017}, i.e. ELM-like timescales, a characteristic frequency of appearance and an ELM-like divertor signature. In addition, they cause a full relaxation of edge profiles, observed in both the electron temperature and electron density (no statement can be made with regard to the ion temperature due to diagnostic time resolution restrictions). For these reasons, we will hereinafter refer to them as I-mode pedestal relaxation events (\PRE{s}). 
\newline In this work, we carry out an extensive investigation of the I-mode pedestal relaxation events that have recently been observed in AUG, shedding light on their appearance, characteristics and divertor heat loads.
In section 2, an illustrative I-mode discharge with two phases -- one with and another without \PRE{s} -- is shown. The appearance of I-mode \PRE{s} in the typical AUG I-mode operational space ($I_p$\,=\,1 MA, $B_{\mathrm{t}}$\,=\,$- 2.5$ T) is also discussed. In section 3, the temporal evolution of the edge profiles during I-mode \PRE{s} is presented, along with their associated energy and particle losses. This section also features a MHD stability analysis of edge profiles during these events. 
A broad analysis of scrape-off layer (SOL) transport during \PRE{s} is carried out in section 4. In particular, the characteristic SOL temporal evolution, filament propagation and their toroidal asymmetric origin are discussed. In section 5, divertor transient heat loads induced by \PRE{s} are investigated, with a focus on energy deposition asymmetries between inner and outer divertor targets and divertor peak energy fluence measurements. In this section, projections to ARC~\cite{Sorbom_2015} are also drawn. 
Lastly, in section 6, the main conclusions are outlined.

\section{Appearance in the typical AUG I-mode parameter range}

\begin{figure*}[htb]
        \centerline{\includegraphics[width=0.69 \textwidth]{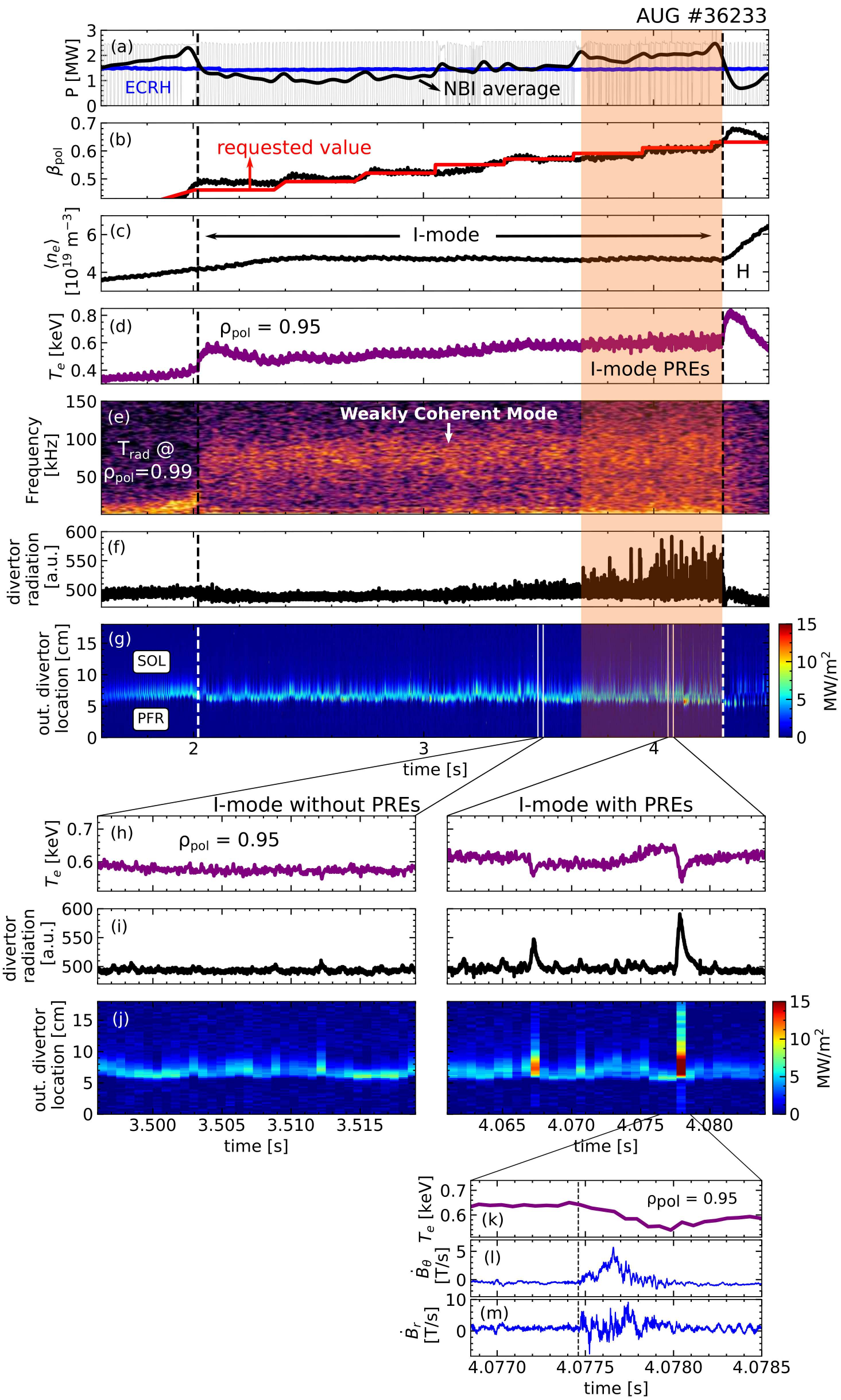}}
        \caption[]{Appearance of I-mode \PRE{s} in the typical AUG parameter range. (a) ECR (blue) and NBI (black) heating power. (b) Requested $\beta_{\mathrm{pol}}$ (red) and measured value (black). (c) Line-averaged core electron density. (d) Electron temperature at $\rho_{\mathrm{pol}}$\,=\,0.95. (e) Spectrogram of ECE radiation temperature measured at $\rho_{\mathrm{pol}}$\,=\,0.99. (f) Radiation measured by a diode bolometer with line of sight in the upper divertor region. (g) Heat flux onto the upper outer divertor target. Panels (h), (i) and (j) show a magnification of the pedestal top electron temperature, divertor radiation and outer divertor heat flux, respectively, in two different time windows with and without \PRE{s}. Panels (k), (l) and (m) show a magnification of the pedestal top electron temperature, the $\dot{B}_{\theta}$ and $\dot{B}_{r}$ signal measured by magnetic pick-up coils during an I-mode \PRE{}.}
        \label{36233}
 \end{figure*}

I-mode plasmas are commonly achieved by using magnetic configurations with the ion $ \nabla B $ drift pointing away from the active X-point, i.e. the so-called unfavorable configuration in terms of H-mode access. In AUG, plasma discharges in the unfavorable configuration are more frequently performed in the upper single null (USN) configuration. This is because the upper divertor tiles are not toroidally tilted, allowing more flexibility in the magnetic field line direction (see e.g.~\cite{Eich_2005ppcf}).
Figure~\ref{36233} shows an I-mode USN discharge where \PRE{s} 
  \begin{figure*}[hbt]
        \centerline{\includegraphics[width=0.95 \textwidth]{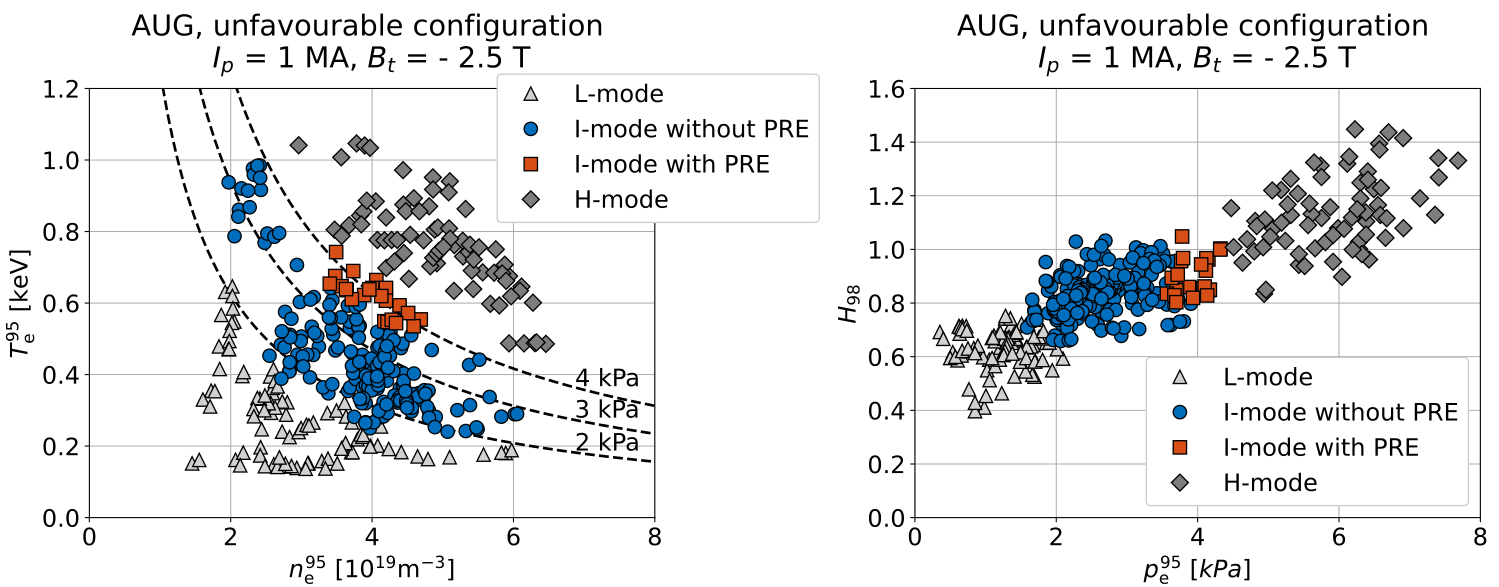}}
        \caption[]{Left: Electron temperature vs. electron density (both evaluated at $\rho_{\mathrm{pol}}$\,=\,0.95) of AUG L-mode, I-mode and H-mode discharges at $I_{p}$\,=\,1\,MA and $B_{t}$\,=\,$-2.5$\,T. For $n_e$\,$\approx$\,4\,$ \times$\,$ 10^{19}$\,m$^{-3}$, I-mode \PRE{s} appear just before the I-H transition. Dashed lines indicate isobars. Right: $H_{98}$ against electron pressure at $\rho_{\mathrm{pol}}$\,=\,0.95. I-mode discharges with and without \PRE{s} can reach $H_{98}$\,=\,1.}
        \label{tn_diag}
 \end{figure*} 
 \noindent occur. This plasma is in the typical AUG I-mode parameter range, i.e. $I_p$\,=\,1\,MA, $B_{\mathrm{t}}$\,=\,$- 2.5$\,T and with a line-averaged core electron density of about $5 \times 10^{19} \mathrm{m}^{-3}$ (see panel\,(c)). The plasma is heated via a constant 1.5\,MW of electron cyclotron resonance heating (ECRH) power and (modulated) co-current neutral beam injection (NBI) power (see panel\,(a)). The NBI power is ramped linearly until $t$\,=\,2\,s when the NBI feedback control on the $\beta_{\mathrm{pol}}$ value takes over (panel\,(b)), similarly to~\cite{Happel_2019}. The $\beta_{\mathrm{pol}}$ is defined as $\overline{p}/(B_{\mathrm{pol}}^{2}/2\mu_0)$, where $\overline{p}$ is the average plasma pressure and $B_{\mathrm{pol}}$ is the average poloidal magnetic field strength. At $t$\,=\,2.02\,s the transition from L to I-mode occurs, which is seen from the sudden increase in the electron temperature at $\rho_{\mathrm{pol}}$\,=\,0.95 (panel\,(d)) and from the appearance of the weakly coherent mode (WCM) in the spectrogram of the radiation temperature (panel\,(e)) measured at $\rho_{\mathrm{pol}}$\,=\,0.99 from electron cyclotron emission (ECE). Simultaneously, as the measured $\beta_{\mathrm{pol}}$ becomes larger than the requested one, the NBI power is reduced to match the two values. This reduction in the NBI power is crucial for achieving stationary NBI-heated I-mode plasmas in AUG, as has already been shown in~\cite{Happel_2019}. Later in the discharge, the request value of $\beta_{\mathrm{pol}}$ is increased in a stepwise manner; both $\beta_{\mathrm{pol}}$ and the pedestal top electron temperature rise accordingly, while the electron density stays constant. 
 At $t$\,=\,4.3\,s the transition to H-mode takes place, which can be seen from the simultaneous rise of both the pedestal top electron temperature and the line-averaged electron density. Also, the WCM disappears from the edge electron temperature spectrogram when the H-mode begins. In AUG, I-mode \PRE{s} appear before the I-H transition (in this case when $\beta_{\mathrm{pol}}$\,$\approx $\,0.58\,$ -$\,0.61). This can be seen by looking at the radiation measured with an AXUV diode bolometer~\cite{Bernert_2014} having a line of sight in the upper divertor region (panel\,(f)). The radiation is characterized by several spikes in this phase and the heat flux onto the upper outer divertor inferred from infrared cameras~\cite{Sieglin_2015} exhibits transient rises (panel\,(g)). A clearer view of I-mode \PRE{s} is shown in the magnification in the bottom right panel of Fig.~\ref{36233}. The pedestal top electron temperature drops during these events (panel\,(h)). The plasma expelled from the confined region enters the SOL and reaches the divertor region, causing a strong increase in the radiation signal (panel\,(i)). Ultimately, it hits the divertor target, leading to a substantial increase in the divertor heat fluxes (panel\,(j)). For comparison, a magnification of the same time traces of an I-mode without \PRE{s} is shown in the bottom left panel of Fig.~\ref{36233}. The pedestal top electron temperature is rather constant and does not exhibit large drops. This is reflected in the divertor radiation and heat flux values that do not change substantially. Additionally, panels\,(l) and (k) show the temporal derivative of the poloidal and radial component of the magnetic field measured by magnetic pick-up coils during an I-mode \PRE{}. After the \PRE{} onset (indicated by a vertical dashed line), both the poloidal and radial component of the magnetic field are perturbed. Moreover, no precursor oscillations are detected by magnetic pick-up coils before the onset of \PRE{s}, marking a difference with type-I, type-II and type-III ELMs, which are characterized by detectable magnetic precursors~\cite{Zohm_1996, Doyle_1991, Stober_2001}.
 \newline Figure~\ref{tn_diag}\,(a) shows the AUG I-mode discharges obtained at $I_{p}$\,=\,1\,MA and $B_t$\,=\,$-2.5$\,T in the $T_e$-$ n_e$ operational space. Also L-mode and H-mode discharges obtained at the same plasma current and toroidal magnetic field are plotted. Electron temperature and density at $\rho_{\mathrm{pol}}$\,=\,0.95 have been obtained through integrated data analysis (IDA)~\cite{Fischer_2010}, which combines different diagnostics measurements (such as edge and core Thomson scattering, electron cyclotron emission radiometers and DCN interferometry). I-mode with \PRE{s} appear only in a small portion of the $T_e$-$n_e$ diagram, namely around the isobar at 4\,kPa for 3\,$\times$\,$10^{19}$\,$<$\,$n_e^{95}$\,$<$\,5\,$ \times$\,$ 10^{19}$\,m$^{-3}$. This isobar lies very close to the I-H transition. Figure~\ref{tn_diag}\,(b) shows the same database plotted in the $H_{98}$-$p_e^{\mathrm{95}}$ space, where $H_{98}$ is the energy confinement time normalized to the IPB98(y,2) scaling law~\cite{Transport_1999} and $p_e^{\mathrm{95}}$ is the electron pressure at $\rho_{\mathrm{pol}}$\,=\,0.95. When I-mode \PRE{s} appear, no substantial enhancement of $H_{98}$ is observed with increasing $p_e^{\mathrm{95}}$, i.e. I-mode discharges with and without \PRE{s} can reach high $H_{98}$ ($ \simeq$\,0.8\,$-$\,1.0). This indicates that in AUG stationary and high-confinement I-mode plasmas without \PRE{s} can be regularly achieved.
 At this stage, it is not clear whether \PRE{s} are present in I-mode at higher densities. At very low densities, I-mode \PRE{s} have not been found at 1\,MA/2.5\,T. Density and current dependence of I-mode \PRE{s} will be subject of future research at AUG.

\section{Edge profiles evolution, losses and stability}
\subsection{Temporal evolution}

\begin{figure}[htb]
        \centerline{\includegraphics[width=0.5 \textwidth]{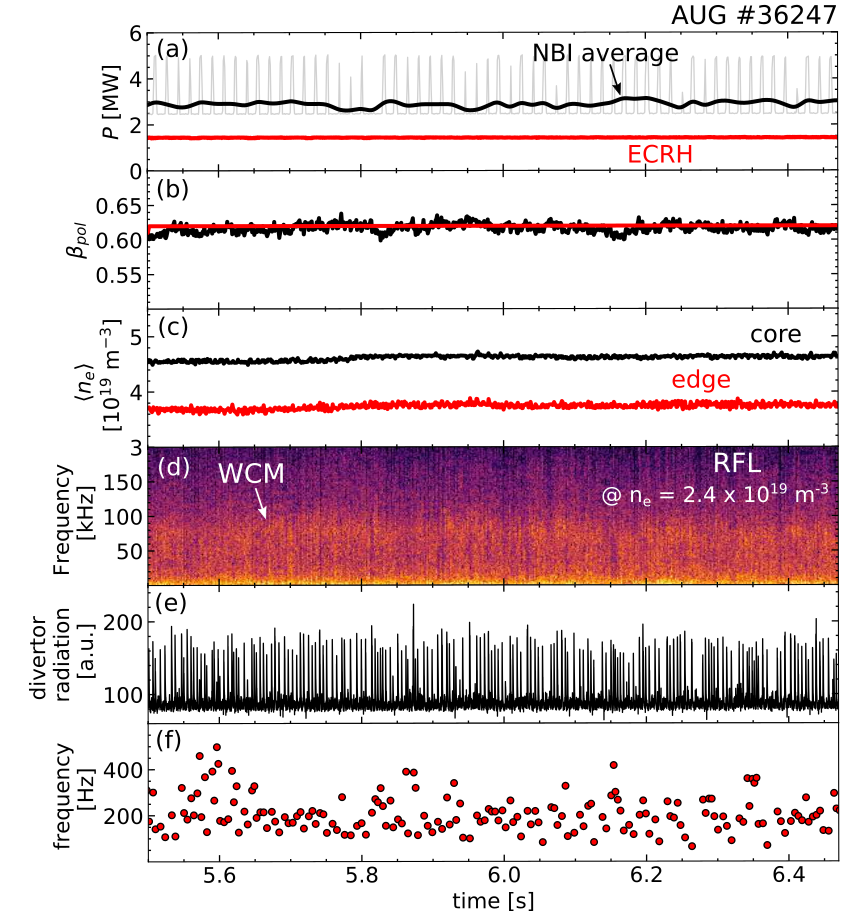}}
        \caption[]{Stationary I-mode discharge with \PRE{s}. (a) ECRH (red) and NBI (gray) heating power. The solid black line is the averaged NBI power. (b) Requested $\beta_{\mathrm{pol}}$ (red) and measured value (black). (c) Line-averaged core (black) and edge (red) electron densities. (d) O-mode reflectometry homodyne signal spectrogram caused by density fluctuations at $n_{\mathrm{e}}$\,=\,2.4\,$\times$\,$10^{19}$\,$ \mathrm{m}^{-3}$, i.e. $\rho_{\mathrm{pol}}$\,$\simeq$\,0.98. (e) Radiation in the upper divertor region. (f) Frequency of \PRE{s}.}
        \label{36247}
\end{figure}

Thanks to the $\beta_{\mathrm{pol}}$ NBI feedback control, stationary I-mode discharges with several \PRE{s} have been obtained. Figure~\ref{36247} shows the time trace of different quantities during such an I-mode discharge (an USN plasma with $I_p$\,=\,1\,MA and $B_{\mathrm{t}}$\,=\,$-2.5$\,T, see e.g.~\cite{Happel_2017}). The heating power (panel\,(a)) is a mixture of constant ECRH power and modulated NBI power, with the latter feedback controlled on the requested $\beta_{\mathrm{pol}}$ value (in this case 0.61, see panel\,(b)). Both the core and edge line-averaged electron density (panel\,(c)) are constant in this phase. The WCM is present during the entire I-mode and it peaks at around 75\,kHz, as can be seen in the spectrogram of the reflectometry signal~\cite{Silva_1996} detected at $\rho_{\mathrm{pol}}$\,$\simeq$\,0.98 (panel\,(d)), which is a measure of electron density fluctuations. The divertor radiation measured by diode bolometers with line of sight in the upper divertor region is used in this work as a \PRE{} monitor, as it shows a clear peak during each \PRE{} event (panel\,(e)). The frequency of occurrence of I-mode \PRE{s} ranges between 100 and 400\,Hz (panel\,(f)) in this discharge. 
\newline The constant plasma parameters and \PRE{} behavior found in this I-mode discharge allowed us to obtain the conditionally averaged temporal evolution of edge electron temperature and density profiles during \PRE{s}. Fast measurements of electron density were obtained from DCN interferometry~\cite{Mlynek_2010} and lithium beam emission spectroscopy (Li-BES)~\cite{Schweinzer_1992, Willensdorfer_2014}, while the electron temperature was measured with electron cyclotron emission (ECE) radiometers~\cite{Denk_2018}. These diagnostics measurements were combined through integrated data analysis (IDA)~\cite{Fischer_2010}, which reconstructs electron temperature and density profiles in the framework of Bayesian probability theory with a time resolution of 0.1\,ms. The reconstructed IDA temperature profiles have separatrix temperatures around 60\,$-$\,80\,eV, which is the typical I-mode separatrix temperature at AUG measured by the edge Thomson scattering (TS) system~\cite{Kurzan2011}. Similar separatrix temperatures are also obtained applying the two-point model~\cite{Silvagni_2020}. 
In this work, Li-BES data during each \PRE{} were corrected by subtracting the enhanced passive radiation observed during the \PRE{} phase, with the same method used in~\cite{Cavedon_2017} to analyze the type-I ELM cycle. For details refer to~\cite{Cavedon_2017}. Hereinafter, the time synchronization of each \PRE{} was obtained using a diode bolometer channel with line-of-sight in the upper divertor region, as it has been found to be the best \PRE{} monitor signal in USN plasma discharges. 
\begin{figure}[htb]
        \centerline{\includegraphics[width=0.5 \textwidth]{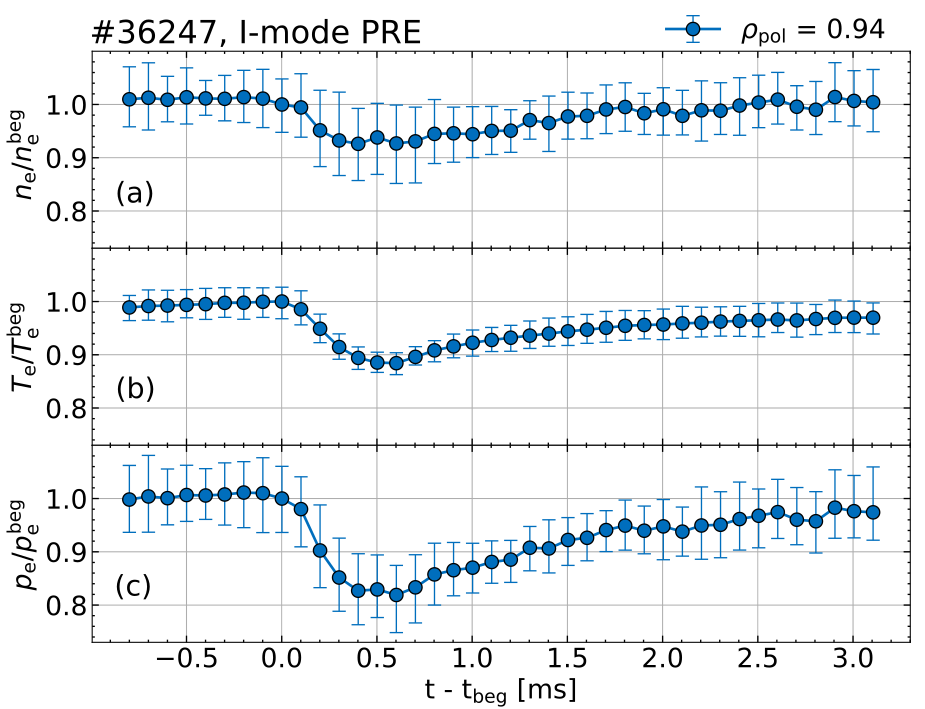}}
        \caption[]{Temporal evolution of conditionally averaged electron density (a) temperature (b) and pressure (c) at $\rho_{\mathrm{pol}}$\,=\,0.94 (i.e. around the electron temperature pedestal top position). Each quantity is normalized to its value at the \PRE{} onset.}
        \label{pedestal_evolution}
\end{figure}
\begin{figure*}[htb]
        \centerline{\includegraphics[width=1 \textwidth]{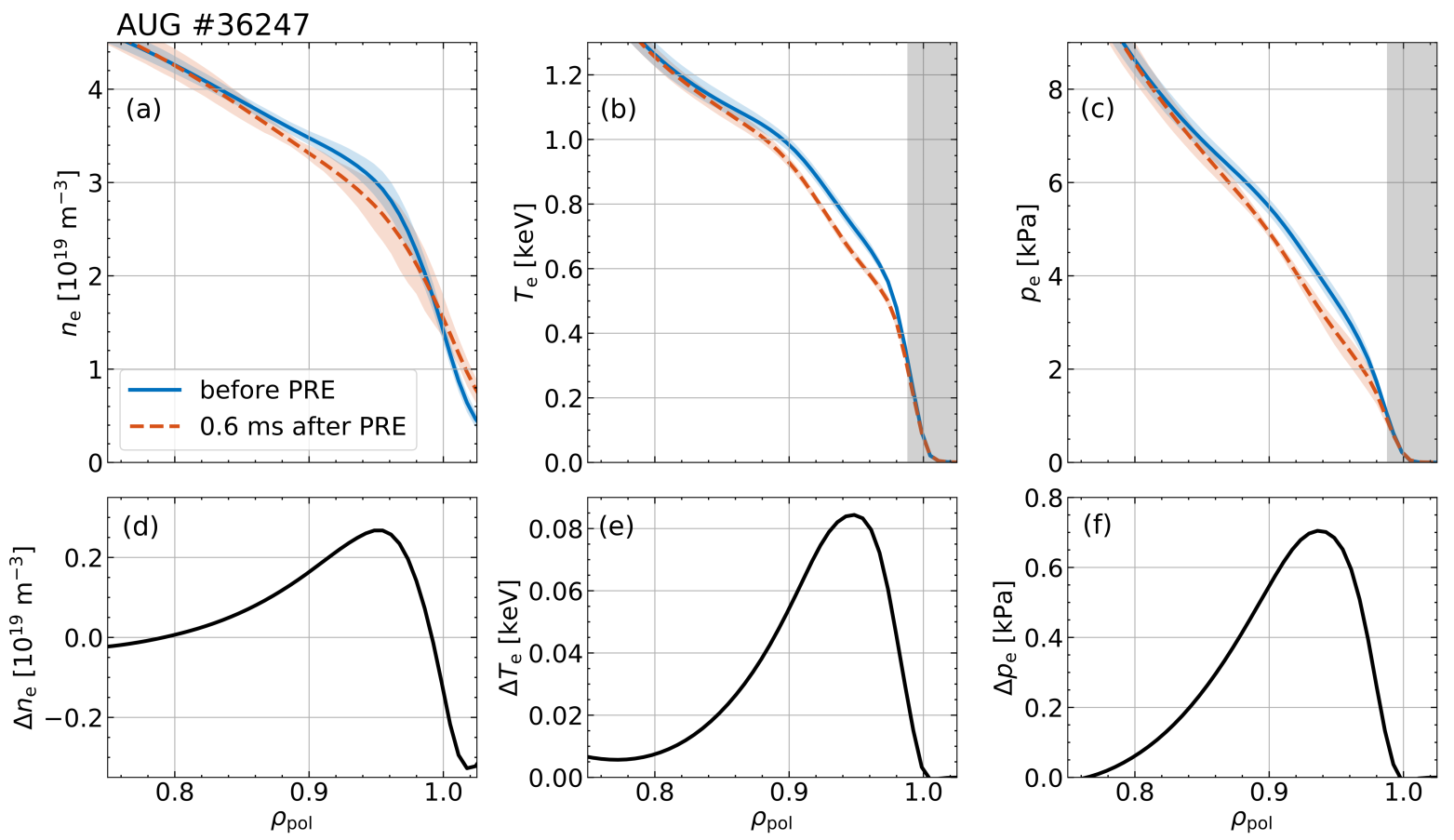}}
        \caption[]{Top: Conditionally averaged radial profiles of electron density (a), temperature (b) and pressure (c) before and 0.6\,ms after I-mode \PRE{s}. The shaded gray area indicates the ECE ``shine-through'' region. Bottom: difference between the before \PRE{} and after \PRE{} radial profiles of the electron density (d), temperature (e) and pressure (f).}
        \label{profile_crash}
 \end{figure*} 
 Then, different plasma quantities were conditionally averaged using the above-mentioned time synchronization in the time window between $t_{\mathrm{start}}$\,=\,5.5\,s and $t_{\mathrm{end}}$\,=\,6.47\,s. Only \PRE{s} occurring when Li-BES was switched on are considered, giving a total number of 105 \PRE{s}. Figure~\ref{pedestal_evolution} shows the time evolution of conditionally averaged electron temperature, density and pressure evaluated around the temperature pedestal top position, i.e. $\rho_{\mathrm{pol}}$\,=\,0.94. Error bars represent the 50$\%$ confidence interval of the 105 events analyzed. All quantities are normalized to the value assumed at the \PRE{} onset. The relative drop of the electron density is about 7\,$\%$, whereas the electron temperature reduces by 12\,$\%$. As a consequence, the electron pressure decreases by about 18\,$\%$ during the \PRE{}. 
The relative drop of the pedestal top electron temperature and density during I-mode \PRE{s} in $\#$36247 is lower than the corresponding reduction during type-I ELMs, which ranges between 20 and 50\,$\%$ (for the electron temperature) and between 10 and 40\,$\%$ (for the electron density) across different devices~\cite{Cavedon_2017,Frassinetti_2015,Viezzer_2016, Wade_2005}. It is also worth noting that the electron temperature and density exhibit different recovery times: 2.5\,ms after the \PRE{} onset, the electron density is back to its value before the \PRE{}, while the electron temperature has not recovered yet. 
The different recovery time of pedestal top electron density and temperature is a typical feature of the type-I ELM cycle, as reported in~\cite{Burckhart_2010, Viezzer_2016, Cavedon_2017} and it has been shown to be related to the different recovery times of the electron temperature and density pedestal gradients during the ELM cycle.
Figure~\ref{profile_crash} shows a comparison of the electron density (panel\,(a)), temperature (panel\,(b)) and pressure (panel\,(c)) radial profiles before and 0.6\,ms after the \PRE{} onset. The envelopes around each profile represent the 50\,$\%$ confidence interval of the averaged profiles. The error envelopes of the electron density profiles are larger at the pedestal top because at that position the lithium beam signal is weaker. After 0.6\,ms the electron density profile shows a decrease in the edge density gradient, with the density for $\rho_{\mathrm{pol}}$\,$<$\,0.98 reducing and the SOL density increasing. This results in the formation of a pivot point around $\rho_{\mathrm{pol}}$\,=\,0.99. The electron density at the pivot point is approximately equal to half of the density at $\rho_{\mathrm{pol}}$\,=\,0.95. Both the aforementioned features are similar to what happens during type-I ELMs in AUG~\cite{Nunes_2004}. The reconstructed IDA electron temperature radial profile during \PRE{} suffers from the so-called ECE ``shine-through'' effect at the plasma edge~\cite{Barrera_2010}, which makes the evaluation of electron temperature edge gradients around the separatrix challenging in this particular case. For this reason the pivot point of the electron temperature profiles cannot be evaluated. Nonetheless, outside the ``shine-through'' region (shaded gray area in Fig~\ref{profile_crash} (b)), ECE data are reliable, as they display very good agreement with TS data. In this region ($\rho_{\mathrm{pol}}$\,$<$\,0.98), the electron temperature shows a clear reduction after the \PRE{} onset, as it is observed for the density. Additionally, the bottom panels of Fig.~\ref{profile_crash} show the radial profiles of the electron density (panel\,(d)), temperature (panel\,(e)) and pressure (panel\,(f)) collapses, defined as the profile difference before and after the \PRE{}. 
The electron density, temperature and pressure collapses show their maximum around $\rho_{\mathrm{pol}}$\,=\,0.95, i.e. around the I-mode electron temperature pedestal top position. From Fig.~\ref{profile_crash} the \PRE{} profile affected depth can be evaluated, i.e. the distance between the pivot point and the inner radial point where no substantial difference between before and after \PRE{} profile is detected. The \PRE{} affected depth is about 0.2 in terms of normalized poloidal flux radius and 10\,cm in terms of radial distance, similar to the ELM affected depth in AUG~\cite{Nunes_2005}, JET~\cite{Beurskens_2009}, D-IIID~\cite{Leonard_2002} and JT-60U~\cite{Chankin_2002}. 

\subsection{Energy and particle losses}
Knowing the radial profiles' temporal evolution, the I-mode \PRE{} energy and particle losses can be calculated, as can the convective and conductive components of the energy loss. The \PRE{} energy and particle losses are defined as:

 \begin{equation}
  \Delta W_{\mathrm{PRE}}  =  \frac{3}{2} \int \Delta p\, \mathrm{d}V 
  \label{eq:energy_loss}
\end{equation}

 \begin{equation}
  \Delta N_{\mathrm{PRE}}  =    \int \Delta n_{e}\, \mathrm{d}V.
  \label{eq:particle_loss}
\end{equation}
Energy losses are normalized either to the total energy content of the plasma $W_{\mathrm{MHD}}$ (which is evaluated from the reconstructed MHD equilibrium) or to the pedestal energy content, defined as $W_{\mathrm{ped}}$\,=\,$3/2$\,$p_{\mathrm{ped}}V_{\mathrm{plasma}} $. Particle losses are usually normalized to the pedestal particle content, defined as $N_{\mathrm{ped}}$\,=\,$n_{\mathrm{e,ped}}V_{\mathrm{plasma}}$. A previous multi-machine study on type-I ELMs revealed that the relative ELM energy losses scale with the pedestal collisionality~\cite{Loarte_2003}, which is defined as:
\begin{equation}
    \nu^*_{\mathrm{ped}}= R q_{\mathrm{95}} \epsilon^{-3/2} (\lambda_{\mathrm{e,e}})^{-1}
    \label{eq:pedestal_coll}
\end{equation} where $\epsilon$ is the inverse aspect ratio and $\lambda_{\mathrm{e,e}}$\,= \,1.7\,$ \times$\,$10^{17}$\,$T_{\mathrm{e, ped}}^2$\,$(\mathrm{eV}) / [n_{\mathrm{e, ped}}$\,$(\mathrm{m^{-3}}) \mathrm{ln} \Lambda]$ is the electron-electron Coulomb collision mean free path at the pedestal. The Coulomb logarithm $\mathrm{ln} \Lambda$ is evaluated here following the classical formula for electron-electron collisions~\cite{Honda_2013} and using pedestal plasma parameters.
The energy loss can be further broken down into two contributions, termed conductive and convective energy losses ($\Delta W_{\mathrm{cond}}$ and $ \Delta W_{\mathrm{conv}}$), which are defined as:
\begin{gather}
  \nonumber \Delta W_{\mathrm{PRE}}  \approx  \frac{3}{2} k \left[\int n \Delta T \mathrm{d}V + \int T \Delta n \mathrm{d}V \right] = \\
  = \Delta W_{\mathrm{cond}} + \Delta W_{\mathrm{conv}},
  \label{eq:cond_conv}
\end{gather}
where the cross-term is neglected, as it is of second order. Breaking down the energy loss into a conductive and convective term is important, since these two contributions may behave and scale differently to larger devices. Indeed, for type-I ELMs it has been shown that convective ELM energy losses have a weak dependence on pedestal plasma parameters~\cite{Leonard_2002}, while conductive ELM energy losses show clear trends with pedestal quantities such as $\nu^*_{\mathrm{ped}}$.
In this subsection, first estimations of the above-mentioned quantities for I-mode \PRE{s} will be shown.
\newline The \PRE{} particle loss is calculated directly from the measured $\Delta n_e$ shown in Fig.~\ref{profile_crash}\,(d). In the analyzed discharge, the \PRE{} particle loss is $8.6$\,$\times$\,$10^{18}$\,m$^{-3}$, which is about 2\,$\%$ of the pedestal particle content $N_{\mathrm{ped}}$. 
This percentage is close the lower possible values of $\Delta N/N_{\mathrm{ped}}$ during ELMs, which ranges between 2\,$\%$ and 13\,$\%$ in AUG~\cite{Nunes_2005} and JET~\cite{Loarte_2003}.
\newline To calculate the energy losses, the ion contribution also needs to be considered. As no fast measurements of $T_{i}$ and $n_{i}$ were available to temporally resolve \PRE{s}, assumptions need to be introduced. In this low-impurity content discharge, $n_i$\,=\,$ n_e$ is assumed. 
Also, since in I-mode discharges at AUG the pedestal top ion and electron temperature are very close~\cite{Happel_2017}, $T_i$\,=\,$ T_e$ is considered. Moreover, it is also conjectured that $\Delta n_e$\,=\,$ \Delta n_i$ and $\Delta T_e$\,=\,$\Delta T_i$. While the first equality has been proven to hold at different collisionalities for type-I ELMs~\cite{Wade_2005}, the second one can be broken at low collisionalities for type-I ELMs~\cite{Wade_2005}, with $\Delta T_i$\,$<$\,$\Delta T_e$. Therefore, the following energy losses should be regarded as an upper limit. 
\newline Considering the above-mentioned assumptions, the conductive and convective energy losses during I-mode \PRE{} are 4.2\,kJ (3.3\,$\%$ of $W_{\mathrm{ped}}$) and 3.2\,kJ (2.5\,$\%$ of $W_{\mathrm{ped}}$), respectively. For the same pedestal collisionality, ELM conductive and convective relative losses range between 6\,$-$\,11\,$\%$ and 6\,$-$\,18\,$\%$~\cite{Leonard_2002}, respectively, and are thus larger than the I-mode \PRE{s} losses. \newline The \PRE{} energy loss obtained from Eq.~\ref{eq:energy_loss} is 7.2\,kJ, which is 1.8\,$\%$ of $W_{\mathrm{MHD}}$ and 5.7\,$\%$ of $W_{\mathrm{ped}}$. 
\begin{figure}[htb]
        \centerline{\includegraphics[width=0.5 \textwidth]{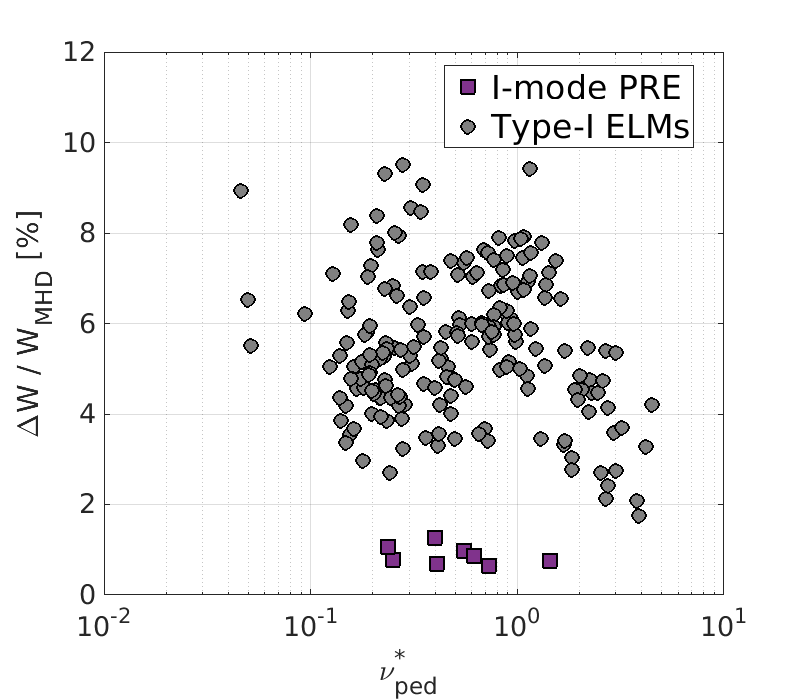}}
        \caption[]{Relative energy loss against pedestal collisionality for type-I ELMs and I-mode \PRE{s}. I-mode \PRE{s} exhibit relative energy losses lower than those of type-I ELMs.}
        \label{enloss_coll}
\end{figure}
This quantity is compared to the energy loss calculated from conditionally averaging the $W_{\mathrm{MHD}}$ signal, which yields $\Delta W_{\mathrm{MHD}}/W_{\mathrm{MHD}}$\,$\approx$\,1.1\,$\%$ and $\Delta W_{\mathrm{MHD}}/W_{\mathrm{ped}} $\,$\approx$\,3.3\,$\%$. This discrepancy may be explained by the assumptions made on the ion contribution, which lead (as expected) to an overestimation of the relative energy loss.
Figure~\ref{enloss_coll} shows the relative energy loss (normalized to $W_{\mathrm{MHD}}$) of I-mode \PRE{s} and type-I ELMs against the pedestal collisionality. I-mode \PRE{} energy losses are calculated carrying out a conditional average of the $W_{\mathrm{MHD}}$ signal during I-mode \PRE{s} for different I-mode discharges. 
I-mode \PRE{} relative energy losses are on average around 1\,$\%$. 
Type-I ELM data used for the multimachine study in~\cite{Eich_2017} are reexamined here. For the same pedestal collisionality, type-I ELM relative energy losses range between 3\,$ -$\,10\,$\%$, and are thus larger than those of I-mode \PRE{s}.

\subsection{MHD stability}
\begin{figure}[htb]
        \centerline{\includegraphics[width=0.5 \textwidth]{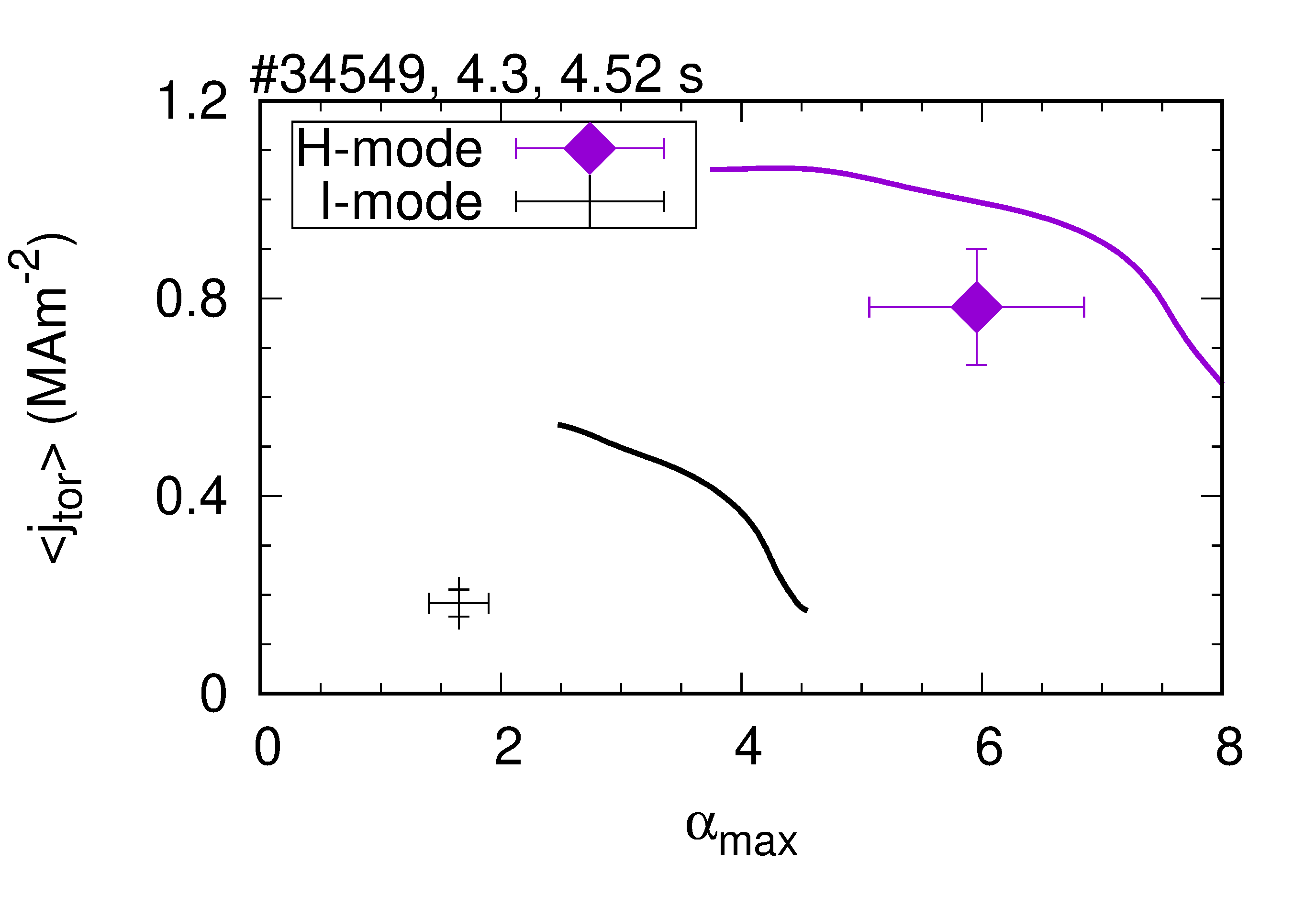}}
        \caption[]{Ideal peeling-ballooning stability diagram of the I-mode pedestal with \PRE{s} (black cross) and the H-mode pedestal with type-I ELMs (purple diamond) in $\#$34549. The black and purple lines represent the stability boundary in I-mode and H-mode, respectively. The I-mode pedestal with \PRE{s} is ideal peeling-ballooning stable.}
        \label{34549_stab}
 \end{figure}

\noindent Figure~\ref{34549_stab} shows the peeling-ballooning pedestal stability diagram calculated with the MISHKA code~\cite{Mikhailovskii_1997} for an AUG discharge with I-mode \PRE{s} that later enters H-mode. The MHD stability is parameterized here in terms of the maximum normalized edge pressure gradient $\alpha_{\mathrm{max}}$~\cite{Connor_1998} and flux-surface averaged edge current density $\langle j_{\mathrm{tor}} \rangle$. The lines show the stability boundary defined with the criterion used in~\cite{Burckhart_2016} for both I-mode and H-mode. Above the stability boundary, peeling-ballooning modes are unstable and type-I ELMs occur. 
The H-mode pedestal is close to the stability boundary and this agrees with the observation of type-I ELMs in the analyzed discharge. On the contrary, the I-mode pedestal with \PRE{s} is far from the ideal peeling-ballooning stability boundary, which is usually the case in I-mode~\cite{Walk_2014, Happel_2017}. This shows clearly that I-mode \PRE{s} are not type-I ELMs. 

\section{Transport in the scrape-off layer}

\subsection{Characteristic temporal evolution}

\begin{figure}[htb]
        \centerline{\includegraphics[width=0.5\textwidth]{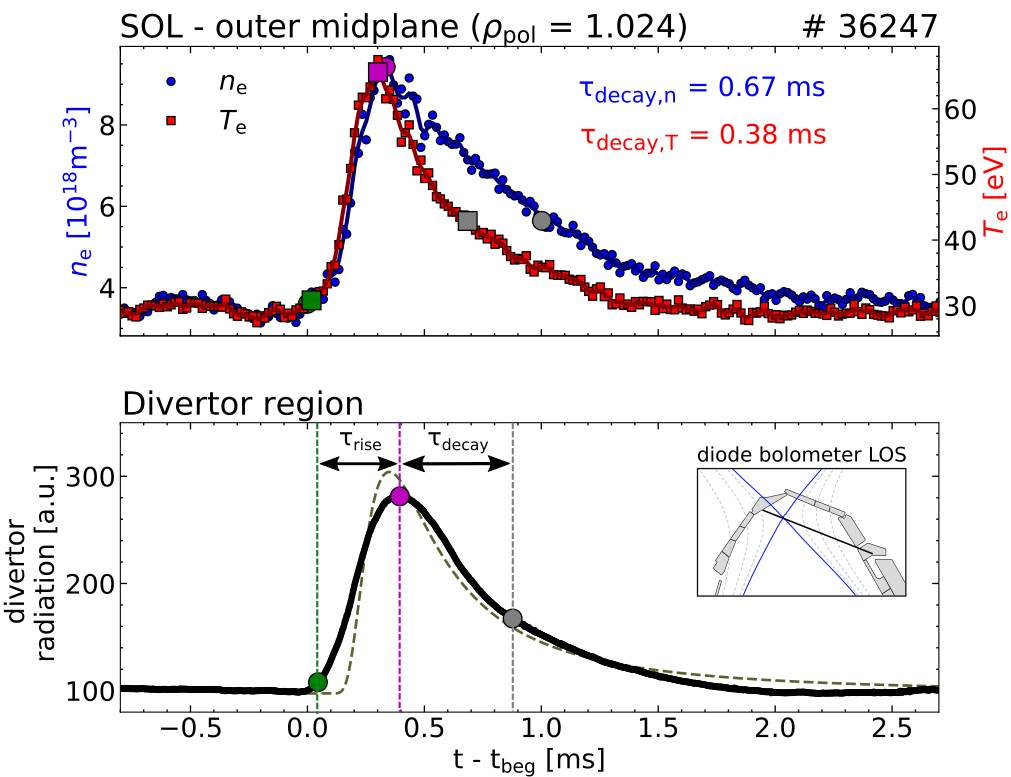}}
        \caption[]{Top: Conditionally averaged temporal evolution of electron temperature (red) and density (blue) at the outer midplane in the SOL ($\rho_{\mathrm{pol}}$\,=\,1.024) measured by the helium beam diagnostics during I-mode \PRE{s}. Bottom: Conditionally averaged divertor radiation during I-mode \PRE{s} measured by a diode bolometer. The line of sight (LOS) of the diode bolometer is depicted in the upper right corner. The dashed line is the fit of the data using Eq.~\ref{eq:fsp}. Green and gray markers represent the beginning and ending time of the \PRE{}, respectively.}
        \label{SOL_div_decays}
 \end{figure}

The energy lost during \PRE{s} from the confined region enters the SOL and is transported both along and across the open magnetic field lines, i.e. both in the parallel and perpendicular direction. The combination of parallel and perpendicular transport in the SOL sets the characteristic temporal evolution of different SOL plasma quantities and of the energy reaching the divertor targets. The temporal shape and duration of the latter quantity is not only important to understand how energy is transported in the SOL, but is also crucial for the assessment of divertor thermal loads~\cite{Sieglin_2017, Yu_2016} and for extrapolation to larger devices. Figure~\ref{SOL_div_decays} shows the conditionally averaged temporal evolution of the outer midplane SOL electron density and temperature (top panel) and of the divertor radiation (bottom panel) during I-mode \PRE{s}. The conditional average was carried out over the same 105 \PRE{s} analyzed in section 3. 
In Fig.~\ref{SOL_div_decays} three time instants are marked: the \PRE{} beginning time $t_{\mathrm{beg}}$ (green), the \PRE{} ending time $t_{\mathrm{end}}$ (gray) and the time instant of the maximum value $t_{\mathrm{max}}$ (magenta). 
In this work, the \PRE{} ending time is defined as the time instant when the peak value decays by $1/\mathrm{e}$, following the definition given in~\cite{Eich_2011} for type-I ELMs. The beginning time is defined as the time instant when the analyzed quantity takes $1/(10\,\mathrm{e})$ of its peak value. From these quantities, the decay time $\tau_{\mathrm{decay}}$\,=\,$ t_{\mathrm{end}}$\,$-$\,$t_{\mathrm{max}}$ and rise time $\tau_{\mathrm{rise}}$\,=\,$ t_{\mathrm{max}}$\,$-$\,$t_{\mathrm{beg}}$ are calculated. The SOL electron density and temperature have been measured by the helium beam system~\cite{Griener_2018} installed at AUG with a time resolution of 17\,$\upmu$s. The radiation is measured in the upper divertor region by a diode bolometer with a time resolution of 2.5\,$\upmu$s. This highly temporally resolved signal is used as a proxy for the energy deposited onto the divertor target plates. IR cameras provide a direct measurement of the energy deposited onto the divertor targets, however their time resolution (about 0.6\,ms) is not high enough to measure the characteristic power temporal shape. The drawback of using the radiation measurements is that the line of sight of the diode bolometer (which is depicted in the bottom panel of Fig.~\ref{SOL_div_decays}) crosses both the outer and inner divertor region. 
Therefore, the radiation temporal evolution shown here should be considered as an ``integral'' of both inner and outer divertor radiation. The top panel of Fig.~\ref{SOL_div_decays} shows that electron temperature and density exhibit a different temporal behavior. While their rise time is almost equal ($\tau_{\mathrm{rise,}n / T}$\,$\approx$\,0.3\,ms), their decay time differs substantially ($\tau_{\mathrm{decay,}T}$\,$=$\,0.38\,ms and $\tau_{\mathrm{decay,}n}$\,$=$\,$0.67$\,ms). This may be due to the fact that conductive transport takes place on a faster timescale than the convective one, i.e. $\chi$\,$\gg$\,$D$, thus giving rise to a faster temperature decay than the density one, as also predicted by the ``free-streaming-particle'' model~\cite{Fundamenski_2005}. The bottom panel of Fig.~\ref{SOL_div_decays} shows that the divertor radiation exhibits the typical temporal evolution of the power deposited onto the divertor targets during type-I ELMs~\cite{Eich_2009}: After an initial sharp rise, a slow exponential decay is present. This characteristic temporal evolution has been successfully modeled in~\cite{Fundamenski_2005} assuming that a Maxwellian distribution of plasma particles (released over a short time $\delta(t)$ compared to the parallel transport times and in a short parallel distribution length $\delta(s)$ compared to the parallel connection lengths to the target) propagates in a ``force-free-way'' along the SOL parallel direction. The heat flux reaching the divertor targets can be thus described by the following equation:
\begin{equation}
    q(t)  =\frac{2 E}{3 \sqrt{\pi}}\left[1+\left(\frac{\tau}{t}\right)^2\right]\frac{\tau}{t^2}\, \mathrm{exp} \left[-\left( \frac{\tau}{t}\right)^2\right] + q_{\mathrm{BG}},  
    \label{eq:fsp}
\end{equation}
\noindent where $E$ is the total energy reaching the divertor, $\tau$ is the characteristic decay time and $q_{\mathrm{BG}}$ is the background heat flux. Equation~\ref{eq:fsp} closely corresponds to the divertor radiation data ($\propto$\,kW/m$^2$) (see dashed curve in the bottom panel of Fig.~\ref{SOL_div_decays}).
Hence, the I-mode \PRE{} power temporal evolution in the SOL does not substantially differ from the type-I ELM one. This points to the universality of energy transport during any pedestal relaxation event.
In addition, Fig.~\ref{SOL_div_decays} shows that the average \PRE{} deposition time (defined as $\tau_{\mathrm{dep}}$\,=\,$ \tau_{\mathrm{rise}}$\,$+$\,$ \tau_{\mathrm{decay}}$) is about 0.8\,ms. This marks a difference with respect to the previously analyzed I-mode density ``bursts'' in AUG, which showed turbulence timescales (of the order of $\upmu$s) both in the confined region and in the SOL~\cite{Happel_2016, Happel_2017}. These longer ``type-I-ELM-like'' timescales reduce the divertor transient thermal load when compared to the shorter turbulence timescales, as they cause a smaller surface temperature rise for the same deposited energy fluence~\cite{Sieglin_2017}.

\subsection{Filament propagation}

 Perpendicular transport in the SOL is dominated by perpendicular advection associated with coherent structures known as filaments. The name comes from the fact that these structures are extended along the magnetic field lines, thereby forming 3D helical filaments~\cite{Kirk_2004, Maqueda_2009, BenAyed_2009}. Filaments carry particle and energy towards the plasma facing components of the main chamber and the divertor target plates~\cite{Eich_PRL2003, Eich_2005ppcf, Devaux_2011}, and are thus a concern for next-generation tokamaks. 
 \begin{figure}[htb]
        \centerline{\includegraphics[width=0.5 \textwidth]{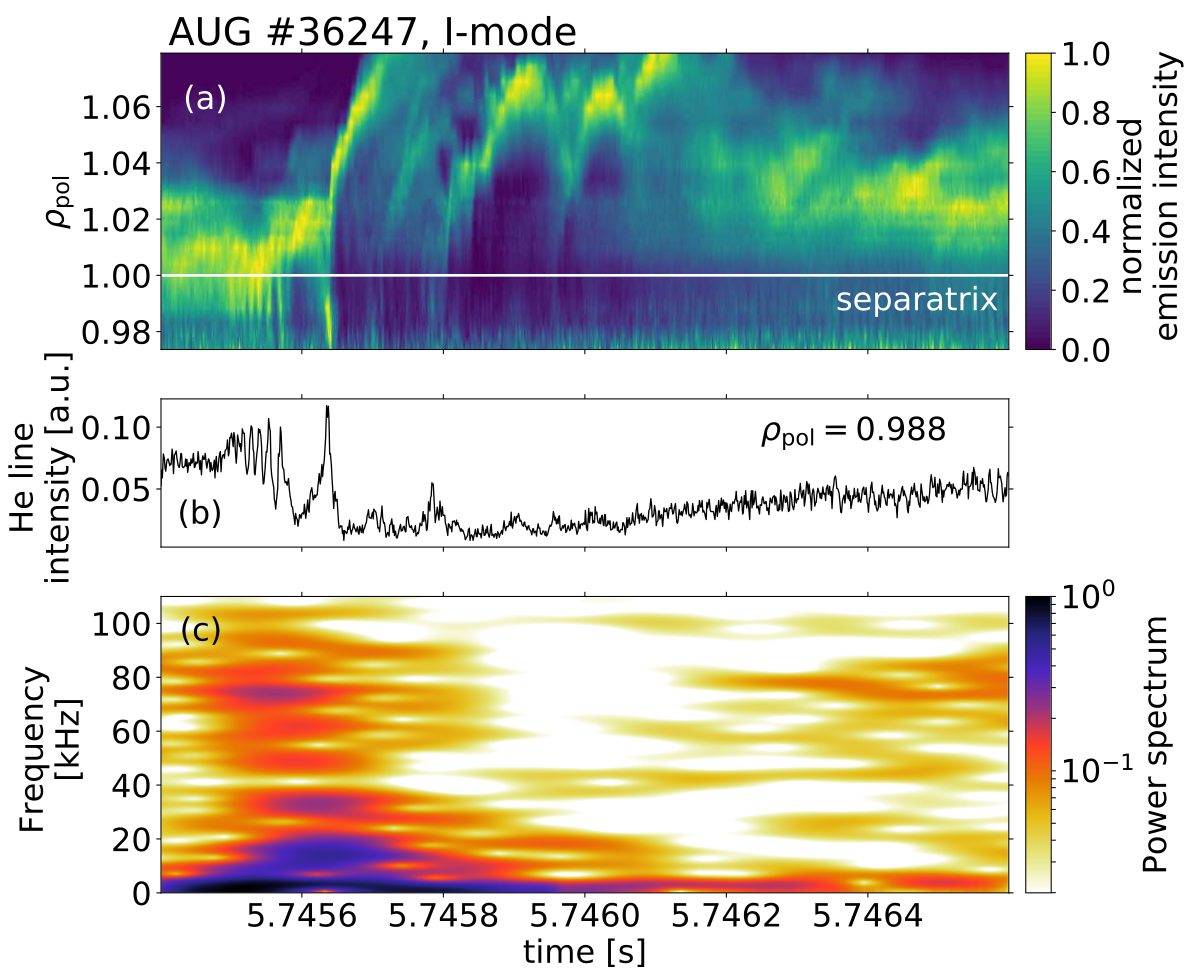}}
        \caption[]{(a) Helium 587.6\,nm line intensity measured at different $\rho_{\mathrm{pol}}$ at the outer midplane by the thermal helium beam system during an I-mode \PRE{}. The line intensity at each $\rho_{\mathrm{pol}}$ has been normalized to its maximum value. Plasma filaments are generated and propagate in the SOL during I-mode \PRE{s}. (b) Helium 587.6\,nm line intensity at $\rho_{\mathrm{pol}}$\,=\,0.988 during an I-mode \PRE{} and (c) its spectrogram. A precursor oscillation at the WCM frequency and localized in the confined region is observed before the onset of this \PRE{}.}
        \label{HEC}
 \end{figure}
 During I-mode \PRE{s}, filaments are generated in the plasma edge and travel through the SOL until reaching the divertor target plates, as is the case for type-I ELMs. The formation and propagation of several filaments during an I-mode \PRE{} is shown in Fig.~\ref{HEC}\,(a). The time evolution of the He 587.6\,nm line intensity measured at different $\rho_{\mathrm{pol}}$ at the outer midplane by the new thermal helium beam system recently installed at AUG~\cite{Griener_2018} is depicted. 
\begin{figure*}[htb]
       \centerline{\includegraphics[width=1 \textwidth]{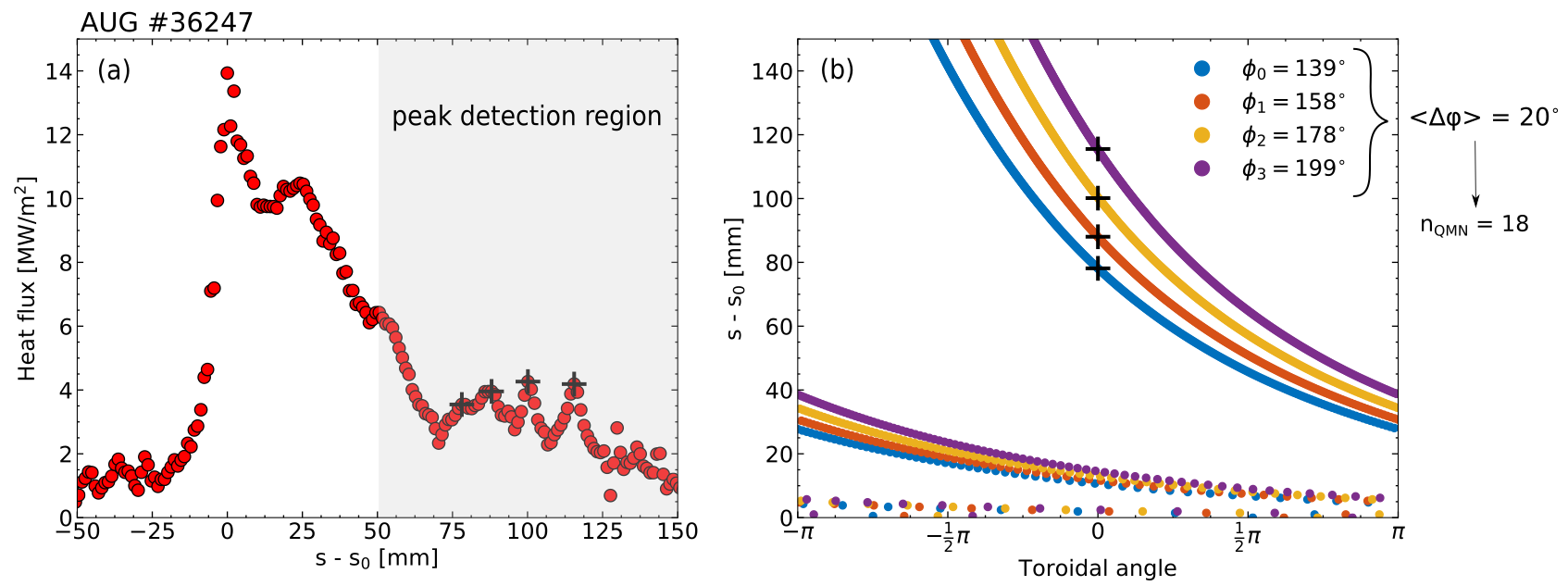}}
       \caption[]{(a) Detected peaks of a heat flux profile during a \PRE{}. The gray region indicates the far-SOL region where the peak detection is carried out. (b) Intersecting structures of the field lines originating in the outer midplane (at different toroidal angles $\phi$) on the upper outer divertor targets. The toroidal angle of the IR tile is set to zero, as a reference. The starting torodial angle of the field lines at the outer midplane is chosen to intersect the location of each peak measured on the outer divertor target.}
       \label{flt}
\end{figure*}
\noindent The line intensity of each channel is normalized to its maximum value. After the onset of the \PRE{} ($t$\,$>$\,5.7456\,s), plasma filaments are expelled into the SOL and propagate both in the radial and toroidal direction. Viewed from a fixed toroidal-poloidal location (e.g. from the outer midplane He-beam viewpoint), the combined effect of toroidal and radial propagation should result in a succession of coherent structures moving radially outwards over time. Indeed, this is what is measured by the helium beam diagnostics, as Fig.~\ref{HEC}\,(a) shows: The plasma filaments propagate radially outwards during an I-mode \PRE{} with a radial velocity distribution function that peaks around 0.5\,km/s. This radial velocity is also similar to the typical filament radial velocities measured during type-I ELMs (0.5\,$ -$\,1\,km/s)~\cite{Kirk_2005, Goncalves_2003, Schmid_2008}, while it is larger than the typical filament radial velocity measured in L-mode, in the inter-ELM H-mode phase and during the I-phase bursts (0.2\,$ -$\,0.3\,km/s)~\cite{Fuchert_2014, Birkenmeier_2014, Griener_2020}. Plasma filaments are generated throughout the entire duration of the \PRE{} (of about 0.7\,ms) until the pedestal recovers. Figure~\ref{HEC}\,(a) shows also that before the \PRE{} onset, a precursor oscillation -- localized in the confined region of the plasma -- is observed. This precursor can be seen more clearly in fig~\ref{HEC}\,(b), where the time evolution of the He 587.6\,nm line intensity measured at $\rho_{\mathrm{pol}}$\,=\,$0.988$ is depicted. The spectrogram of this He line (shown in panel\,(c)) illustrates that the precursor oscillates around $f_{\mathrm{prec}}$\,=\,$75$\,kHz, which is the WCM frequency in discharge $\#$36247, as Fig.~\ref{36247}\,(d) shows. However, it should be mentioned that these precursors in the He 587.6\,nm line intensity are not always observed before the \PRE{} onset. 
Also, no precursors in the $\dot{B_{\mathrm{r}}}$ and $\dot{B_{\mathrm{\theta}}}$ signals measured by magnetic pick-up coils were identified before the onset of I-mode \PRE{s} (see Fig.~\ref{36233}\,(m)). The visibility of a precursor in a local edge diagnostic and the absence of a detectable magnetic precursor from the pick-up coils has been also observed during limit cycle oscillations present in the early I-phase~\cite{Birkenmeier_2016}.

 \subsection{Filament toroidal asymmetric origin}
 
 The filaments observed at the outer midplane ultimately reach the divertor target and are measured via the IR camera system installed at AUG~\cite{Sieglin_2015}. IR cameras measure the divertor surface temperature, and from this, heat fluxes are inferred by using the implicit version~\cite{Nille2018} of the THEODOR code~\cite{Sieglin_2015}, which solves the heat diffusion equation within the divertor tile. Figure~\ref{flt}\,(a) shows an example of the heat flux profile measured on the upper divertor during an I-mode \PRE{}. The peak heat flux is usually observed around the strike point location ($s_{0}$); However, several additional peaks are measured in the far SOL. These divertor substructures are the result of plasma filaments that are generated at the midplane at different torodial locations and that propagate in the SOL in the parallel B-field direction until they hit the divertor targets. The same phenomenon has been observed during type-I ELMs in several devices~\cite{Eich_2005ppcf, Devaux_2011, Pamela_2013}. Magnetic field line mapping from the outer midplane to the divertor target can be used to identify the toroidal displacement of the filaments generated upstream. The procedure is the following: First, the location of the peaks in the far-SOL region is detected as shown in Fig.~\ref{flt}\,(a). Then, the magnetic field lines present at the outer midplane at different radial positions (from the separatrix to the far-SOL) but at the same torodial position are mapped to the upper divertor target. Figure~\ref{flt}\,(b) shows the intersection of such field-line tracing with the upper outer divertor tiles at different torodial angles. The toroidal angle of the IR divertor tile is set to zero. Each coloured line represents the field-line tracing originating from a different toroidal angle at the outer midplane. The starting torodial angle of the field lines at the outer midplane is chosen to intersect the position of each peak measured on the outer divertor target. In the case shown in Fig.~\ref{flt}\,(b), the toroidal angles of the field lines at the outer midplane that are intersecting the peaks are $\phi_0$\,=\,$139^\circ$, $\phi_1$\,=\,$ 158^\circ$, $\phi_2$\,=\,$178^\circ$, $\phi_3$\,=\,$199^\circ$. Since we have identified a subset of several toroidally displaced origins of energy release in the midplane, a ``quasi mode number''
 \begin{equation}
n_{\mathrm{QMN}}  =\frac{1}{a} \sum_j^a \frac{2 \pi}{ \phi_j - \phi_{j+1}}  
 \label{eq:qmn}
\end{equation}
 can be calculated~\cite{Eich_2005ppcf}, where $a$ is the total number of peaks detected, and $j$ is the index indicating each peak. In the example shown in Fig.~\ref{flt}\,(b), $n_{\mathrm{QMN}}$\,=\,18 is obtained. It should be noted that the peak detection is carried out only in the far SOL, i.e. for $s$\,$-$\,$s_0$\,$>$\,50\,mm (which corresponds to $r$\,$-$\,$r_{\mathrm{sep}}$\,$ >$\,7\,mm at the outer midplane). This is because, for $s$\,$-$\,$s_0$\,$<$\,50\,mm, toroidally displaced energy effluxes from the midplane are not sufficiently spaced out at the divertor target, as can be seen from Fig.~\ref{flt}\,(b), making it difficult to identify a univocal correspondence between the peaks measured at the divertor and the toroidal location at the midplane. For the I-mode \PRE{s} analyzed in discharge $\#$36247, the quasi mode number ranges between 10 and 28, and has a mean value of $\langle n_{\mathrm{QMN}}\rangle = 20$.
 
 \section{Divertor transient heat loads}

The \PRE{} energy expelled in the SOL ultimately reaches the divertor target plates causing a transient increase of the divertor surface temperature. A critical parameter for the evaluation of transient thermal divertor loads is the so-called ``heat impact factor'', which for a rectangular heat pulse can be written as:
\begin{equation}
  \Delta T  \propto  \frac{E}{A_{\mathrm{wet}} \sqrt{\tau_{\mathrm{dep}}}} = \frac{\epsilon}{ \sqrt{\tau_{\mathrm{dep}}}},  
  \label{eq:impact_factor}
\end{equation}
where $\Delta T$ is the divertor surface temperature rise during the transient event, $E$ is the deposited energy, $A_{\mathrm{wet}}$ is the wetted area, $\tau_{\mathrm{dep}}$ is the deposition time and $\epsilon$ is the deposited energy fluence (J/m$^2$). Measurements and extrapolation of the above-mentioned quantities to larger devices is of crucial importance to assess the thermal load impact of I-mode \PRE{s} and their divertor compatibility in a fusion power plant. Hereinafter, we present first measurements of the deposited energy and energy fluence onto the inner and outer divertor targets. Some initial considerations regarding \PRE{} compatibility with larger fusion devices are also made.

 \subsection{Energy in-out asymmetry}
   \begin{figure*}[htb]
        \centerline{\includegraphics[width=1 \textwidth]{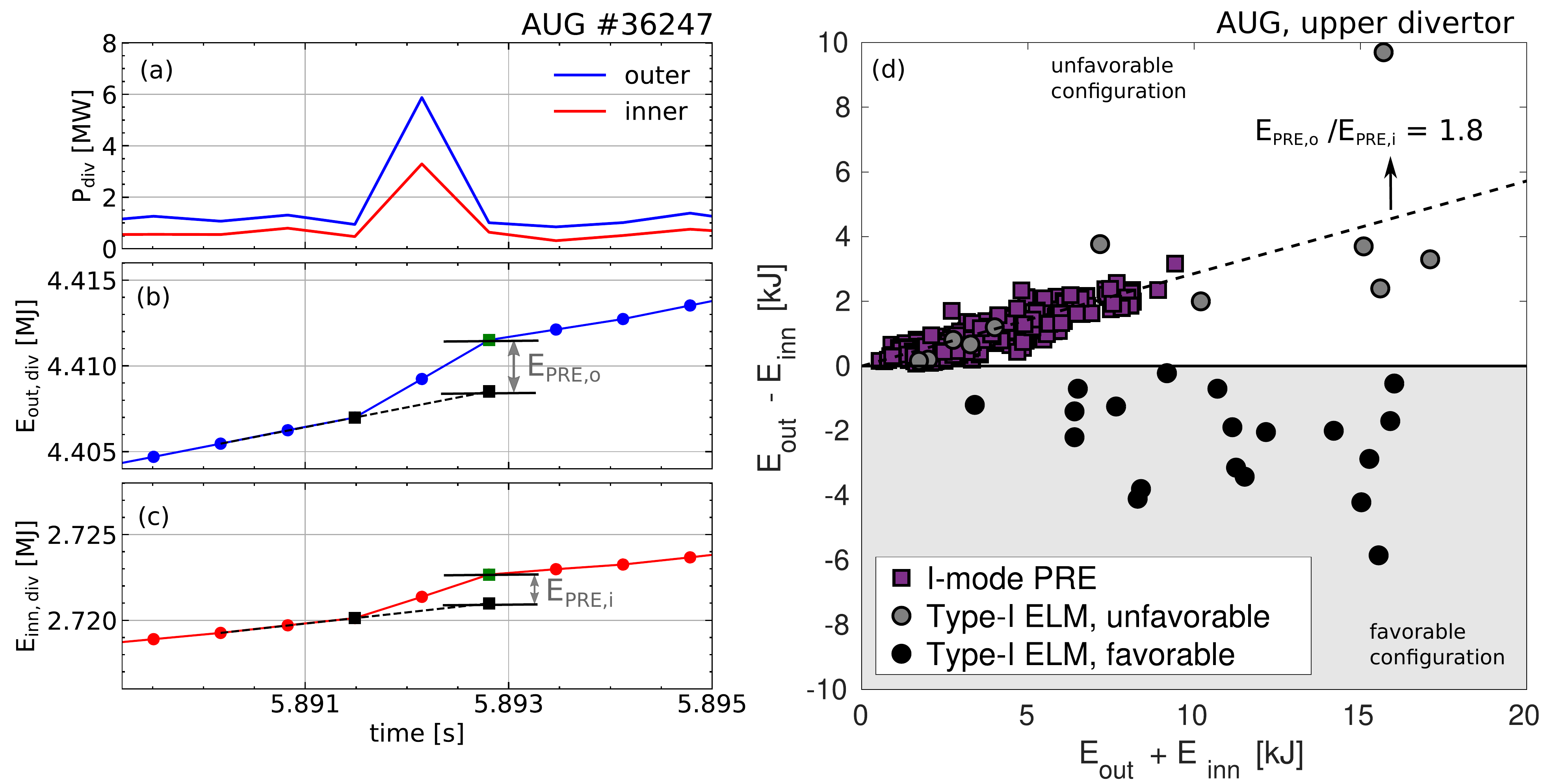}}
        \caption[]{Temporal evolution of outer (blue) and inner (red) power to the upper divertor (panel (a)) and of the cumulative energy reaching the outer (panel (b)) and inner (panel (c)) divertor targets. (d) Difference between the outer and inner divertor energy against the total energy reaching the divertor targets for both I-mode \PRE{s} and type-I ELMs. A clear asymmetry in favor of the outer divertor target is observed for I-mode \PRE{s} and for type-I ELMs in the unfavorable configuration.}
        \label{innout}
 \end{figure*}
The energy expelled from the confined region in the SOL can be transported asymmetrically to the inner and outer divertor targets, leading to the identification of the divertor targets which are more vulnerable to damage. In this subsection, the in-out divertor energy asymmetry during I-mode \PRE{s} is described.
The \PRE{} energy deposited onto the divertor can be evaluated directly from heat flux measurements inferred by the IR camera. The heat flux deposited onto the divertor tile changes across the tile in the toroidal direction, as the incident angle between the magnetic field lines and the divertor target changes (due to the geometry of the tile). Therefore, heat fluxes are evaluated at the toroidal position where the ``tile-average'' heat flux is deposited onto the tile. The power to the inner and outer divertor is calculated by integrating the tile-averaged heat flux over the toroidal angle and over the s-coordinate of the tile (which is the mapping of a radial coordinate on the divertor tile). 
Panel\,(a) in Fig.~\ref{innout} shows the temporal evolution of the inner (red) and outer (blue) divertor power during an I-mode \PRE{}. Both inner and outer divertor exhibit a substantial increase in power during the \PRE{}, which is captured in one single time point. This is consistent with the average \PRE{} deposition time evaluated in the last subsection of about 0.8\,ms and the IR camera measurement frequency of 1500\,Hz (i.e., once every 0.66\,ms). Panel (b) and (c) in Fig.~\ref{innout} show the time evolution of the cumulative energy reaching the outer and inner divertor target, respectively. The cumulative energy is defined as:
 \begin{equation}
    E_{\mathrm{x,div}}(t)  =\int_0^{t} P_{\mathrm{x,div}}\,\mathrm{d}t,
    \label{eq:Ecum}
 \end{equation}
 where the subscript ``x'' can refer to the inner or outer divertor target. The cumulative energy rises linearly over time; however during a \PRE{}, an additional amount of energy reaches the divertor targets, causing $E_{\mathrm{x,div}}$ to jump. The energy increase due to the \PRE{}, $E_{\mathrm{PRE}}$, is then calculated as shown in Fig.~\ref{innout}\,(b)\,-\,(c). 
Figure~\ref{innout}\,(d) shows the difference between the outer and inner divertor \PRE{} energy compared to the total \PRE{} energy reaching the divertor targets for several I-mode USN discharges with \PRE{s}. In addition, type-I ELM data obtained from USN H-mode discharges into a carbon wall (published in~\cite{Eich_2007}) are plotted. All the \PRE{s} analyzed exhibit $E_{\mathrm{PRE,o}}$\,$>$\,$E_{\mathrm{PRE,i}}$, i.e. an energy asymmetry in favor of the outer divertor target. The out/in target energy ratio is about $E_{\mathrm{PRE,o}}/E_{\mathrm{PRE,i}}$\,=\,1.8  in the analyzed discharges. This result is in line with the energy asymmetry previously found in type-I ELM and shown again in Fig.~\ref{innout}\,(d): In the favorable configuration, i.e. when the ion $ \nabla B $ drift points towards the active X-point, more energy is deposited onto the inner divertor target~\cite{Herrmann_2003, Eich_2007, Fenstermacher_2003, Itami_1995}, whereas in the unfavorable configuration, i.e. when the ion $ \nabla B $ drift points away from the active X-point, the asymmetry reverses in favor of the outer target~\cite{Eich_2007}. Hence, the SOL transport during I-mode \PRE{s} does not seem to substantially differ from what has been previously found for type-I ELMs.
The practical consequence of this observation is that the outer divertor target is the more vulnerable to damage due to I-mode \PRE{s}.

\subsection{Peak energy fluence}
The energy fluence deposited on the divertor targets is a crucial parameter for assessing the divertor surface temperature rise during transient events. To allow cross-machine comparisons, divertor geometrical effects must be taken into account, hence in this study the energy fluence parallel to the magnetic field lines is considered.
The parallel energy fluence is calculated by integrating the parallel heat flux profile measured at the divertor target over the \PRE{} duration, which is set by the beginning and ending time defined in section 4: 
\begin{equation}
  \epsilon_{||, \mathrm{PRE}}(s)  =  \int_{t_{\mathrm{beg}}}^{t_{\mathrm{end}}} \frac{q_{\perp} - q_0}{\mathrm{sin}(\mathrm{\alpha_{\mathrm{div}}})} \mathrm{d}t,  
  \label{eq:energyfluence}
\end{equation}
where $\alpha_{\mathrm{div}}$ is the angle between the magnetic field lines and the divertor target. It should be noted that the inter-PRE heat flux, $q_0$, is subtracted from the perpendicular heat flux reaching the divertor target, $q_{\perp}$. In this way, an energy fluence profile is obtained for each \PRE{}. Following the pragmatic approach introduced in~\cite{Eich_2017}, only the peak energy fluence is considered here, as this quantity needs to be directly compared to the material limits, and thus will define the allowed operational range. Measurements were carried out in both I-mode USN and LSN discharges. Due to the wide view of the IR camera over the upper open divertor, both inner and outer parallel energy fluences are available in USN I-mode plasmas. In contrast, only outer divertor heat fluxes are available for the LSN I-mode discharges analyzed in this work. The main parameters of the analyzed discharges are summarized in table~\ref{database}. There are no geometrical differences in the analyzed discharges - triangularity $\delta$ and elongation $\kappa$ are constant. 
\begin{table}[htb]
    {\footnotesize \centerline{
    \begin{tabular}{l|cc}
        \hline\hline
         &  \makecell{LSN } & \makecell{USN } \\
        \hline
        Discharges & 4 & 4 \\
        $W_{\mathrm{MHD}}$ (kJ) & 174--242 & 388--397 \\
        $\overline{n}_e$ ($10^{19}$ m$^{-3}$) & 3.3--4.5 & 4.6--5.2  \\
        $p_{\mathrm{e}}^{\mathrm{ped}}$ (kPa) & 1.8--3.1 & 3.5--4.3 \\
        $T_{\mathrm{e}}^{\mathrm{ped}}$ (keV) & 0.4--0.6 & 0.6--0.8 \\
        $q_{95}$ & 5.2--7.2 & 4.1 \\
        $I_{\mathrm{p}}$ (MA) & 0.6--0.8 & 1.0 \\
        $\beta_{\mathrm{pol}}$ & 0.5--0.9 & 0.6 \\
        $B_{\mathrm{t}}$ (T) & 2.5 & 2.5 \\
        $\delta$ & 0.2 & 0.2 \\
        $\kappa$ & 1.7 & 1.7 \\
        \hline\hline
    \end{tabular}}}
    \caption[]{Parameter range of the I-mode ASDEX Upgrade discharges analyzed.}
    \label{database}
\end{table}
Also, no change in the toroidal magnetic field $B_{t}$ is present, while a plasma current $I_{p}$ (and consequently $q_{95}$ and $\beta_{\mathrm{pol}}$) variation is introduced with the LSN discharges. The plasma stored energy $W_{\mathrm{MHD}}$, as well as the line-averaged core electron density $\overline{n}_{e}$, is lower for the LSN discharges than for the USN pulses. 
\begin{figure}[htb]
        \centerline{\includegraphics[width=0.5 \textwidth]{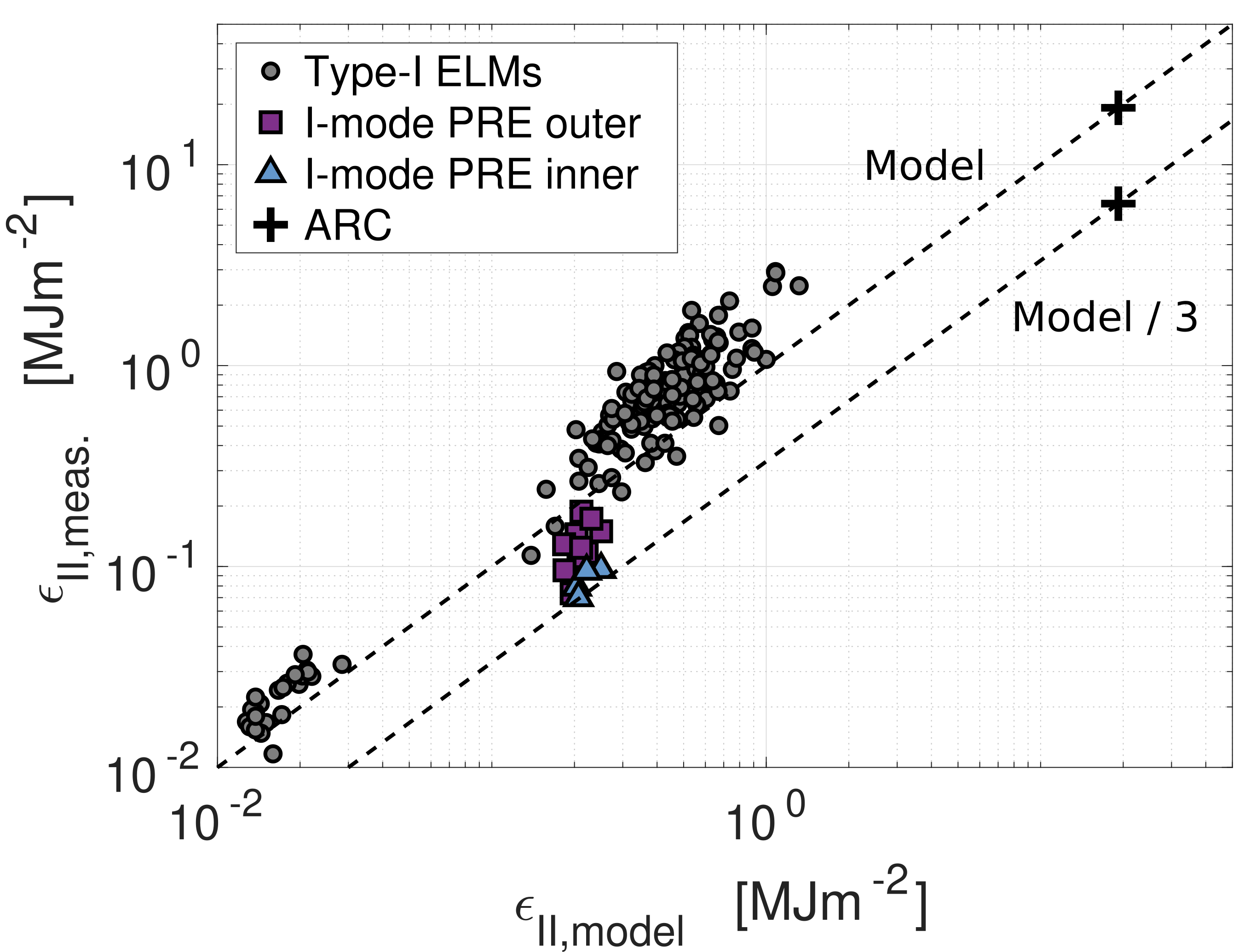}}
        \caption[]{Measured parallel peak energy fluence against the model prediction. The peak energy fluence of I-mode \PRE{s} is lower than that of type-I ELMs for the same $ \epsilon_{||\mathrm{, model}}$. The model prediction gives an upper boundary of the I-mode \PRE{} energy fluences. }
        \label{PRE_scal}
\end{figure}
In addition, a variation in the pedestal top electron pressure and temperature, $p_e^{\mathrm{ped}}$ and $T_e^{\mathrm{ped}}$, is present.
The peak energy fluence shown below is the average value of an ensemble of \PRE{s} taken in phases with constant plasma parameters. 
\noindent These measured values are plotted in Fig.~\ref{PRE_scal} against the parallel energy fluence predicted by the model introduced for type-I ELMs~\cite{Eich_2017}:
\begin{equation}
  \epsilon_{||\mathrm{, model}}  =  \Delta_{\mathrm{equi}} 2 \pi a \sqrt{\frac{1+\kappa^2}{2}} \frac{3}{2} p_{e\mathrm{, ped}} \frac{B_{\mathrm{tor}}^{\mathrm{MP}}}{B_{\mathrm{pol}}^{\mathrm{MP}}}  
  \label{eq:epsmodel}
\end{equation}
where the superscript ``MP'' stands for outer midplane, $a$ is the minor radius, $p_{e\mathrm{, ped}}$ is the pedestal top electron pressure and $\Delta_{\mathrm{equi}}$ is a geometrical factor derived by comparison of the assumed elliptical plasma shape to the real equilibrium reconstruction, which is $\sim $\,2.0 for the ASDEX Upgrade plasma shapes analyzed here. 
In Fig.~\ref{PRE_scal}, type-I ELM peak energy fluence data from several devices~\cite{Eich_2017} are also plotted for comparison. I-mode \PRE{s} exhibit parallel peak energy fluences that are smaller than those of type-I ELMs for the same $\epsilon_{||\mathrm{, model}}$. Also inner divertor peak energy fluences are consistently lower than the outer divertor ones, for the USN discharges analyzed here. In Fig.~\ref{PRE_scal} we additionally draw a line that shows a 0.33-times lower estimate (0.33:1) than the model predicts (1:1). The I-mode \PRE{} energy fluence data fall between these 1:1 and 0.33:1 lines. In particular, Eq.~\ref{eq:epsmodel} represents an upper boundary for I-mode \PRE{}, while it is a lower boundary for type-I ELMs.
The restricted energy fluence \PRE{} dataset available so far does not allow us to conclude that I-mode \PRE{} energy fluences scale similarly to type-I ELMs, although these data show a similar dependency. In particular, in the dataset in question, there is no large variation (when compared to a multi-machine study) of the pedestal top electron pressure. Therefore, achievement and study of I-mode \PRE{s} in other devices is fundamental for any robust extrapolation to future devices. Nonetheless, initial preliminary projections may be attempted assuming that the energy fluence model introduced for type-I ELMs also represents I-mode \PRE{s} well. 
In support of this assumption is not only the overall close correspondence between first I-mode \PRE{} energy fluence measurements and the model, but also the several SOL transport similarities between type-I ELMs and I-mode \PRE{s} shown in this work. 

\subsection{Projections to ARC}
In this subsection, energy fluence projections to ARC~\cite{Sorbom_2015} -- a compact machine ($R$\,=\,3.3\,m and $a$\,=\,1.13\,m) designed to operate in I-mode -- are drawn. The envisaged I-mode pedestal top electron pressure in ARC is around 64\,kPa~\cite{Sorbom_2015} for the $B_0$\,=\,9.2\,T and $q_{95}$\,=\,6 scenario. This leads to a projected parallel peak energy fluence that ranges between 6\,$-$\,19\,MJ/m$^{2}$, with the upper and lower boundaries given by Eq.~\ref{eq:epsmodel} and its 0.33-times lower value.
Divertor material limits strongly depend on the divertor target geometrical design and on the plasma facing component material. The conceptual design of the ARC divertor~\cite{Kuang_2018} has to date been based on the ITER divertor design~\cite{Hirai_2013, Carpentier_2014}, therefore ITER material limits are used in the following discussion. In this respect, a deposited energy fluence limit of about 0.3\,MJ/m$^{2}$ has been found after exposure of ITER divertor monoblocks to 10$^{5}$ cycles of 500\,$\upmu$s-long transient events~\cite{Loewenhoff_2015}. A most recent study that took into account geometrical effects of the castellated structure further lowered this limit to $\epsilon_{\perp, \mathrm{lim}}$\,=\,0.15\,MJ/m$^{2}$~\cite{Gunn_2017}. This limit may be further refined for ARC, given the presence of molten salt coolant instead of water, which needs corrosion issues to be taken into account. Nevertheless, using the actual ITER material limit and an optimistic perpendicular-to-parallel conversion factor of 20, the limit parallel energy fluence at the divertor target is 3\,MJ/m$^{2}$. The lower and higher \PRE{} projected peak energy fluences are a factor of 2 and 6 above the limit,  respectively.
However, it should be noted that the ARC divertor will benefit from the double null configuration that includes a long-leg and a secondary X-point divertor geometry in both the upper and lower divertor chambers~\cite{Kuang_2018}. This configuration will introduce enhanced SOL dissipation effects and a different heat flux distribution on the targets with respect to present-day machines. Therefore, it is expected that the above-mentioned projected values could be lowered by this divertor configuration.
Additional research is needed to provide more detailed analysis of the \PRE{} compatibility with the ARC divertor, among which of primary importance are ad hoc material limits based on the ARC divertor design, multi-machine experimental studies on \PRE{} divertor heat loads and simulations to investigate the beneficial effect of the long-leg secondary X-point divertor geometry w.r.t. present-day single-X point diverted machines.

\section{Conclusions}
This work reports on a first extensive study of I-mode pedestal relaxation events at AUG. It is shown that I-mode \PRE{s} appear close to the H-mode transition in the typical AUG I-mode operational range. Also, no further enhancement of the energy confinement time is observed when \PRE{s} appear, i.e. I-mode discharges both with and without \PRE{s} can reach high $H_{98}$ values ($\simeq$\,0.8\,$ -$\,1.0). This indicates that in AUG stationary and high-confinement I-mode plasmas without \PRE{s} can be achieved. 
\newline Having established the operational range when I-mode \PRE{s} occur in AUG, it was possible for the first time to obtain stationary I-mode discharges with several \PRE{s}, and to study their characteristics. Both I-mode edge electron temperature and density profiles exhibit a relaxation during \PRE{s}. This, in turn, leads to a relaxation of the edge pressure profile, and hence to a loss of energy from the confined region. The edge profile affected depth is about 10\,cm and the largest drop is found around the pedestal top position.
The relative \PRE{} energy loss $\Delta W / W_{\mathrm{MHD}}$ is about 1\,$\%$, being lower than the corresponding type-I ELM loss for the same pedestal collisionality. MHD stability analysis of the I-mode edge profiles shows that they are far from the ideal peeling-ballooning stability boundary, indicating that I-mode \PRE{s} are not type-I ELMs.
\newline In addition, \PRE{} transport in the SOL was investigated. Overall, no major qualitative differences were found between type-I ELM and I-mode \PRE{} SOL transport. They share similar timescales (the deposition time on the divertor is about $\tau_{\mathrm{dep}}^{\mathrm{PRE}}$\,=\,$800$\,$\upmu$s); they both feature the presence of filamentary structures propagating in the SOL, with similar radial velocities ($v_{r}^{\mathrm{PRE}}$\,$ \approx$\,0.5\,km/s); they are both characterized by toroidal asymmetric energy effluxes at the midplane that can be described by a quasi-mode number ($n_{\mathrm{QMN}}^{\mathrm{PRE}}$\,=\,10\,$ -$\,28 with an average value of 20, in the case analyzed); they share the same asymmetry between the inner and outer divertor target deposited energy, namely, in the unfavorable configuration, both I-mode \PRE{s} and type-I ELMs exhibit more energy reaching the outer divertor target. This has the practical consequence that the outer divertor target is the more vulnerable to damage due to I-mode \PRE{s}. 
\newline Lastly, first measurements of the \PRE{} peak parallel energy fluence were shown. They are compared to those of type-I ELMs and to the energy fluence model introduced in~\cite{Eich_2017}. I-mode \PRE{} peak parallel energy fluences are lower than those of type-I ELMs for the same $\epsilon_{||\mathrm{,model}}$. Consequently, the model represents an upper boundary for I-mode \PRE{s}, while a lower boundary is found by dividing the model by a factor of three. Based on these two boundaries, projections to ARC were evaluated, finding $\epsilon_{||}^{\mathrm{peak}}$ between 6\,$-$\,19\,MJ/m$^{2}$, which are above the material limit for the ITER divertor.

\ack
The authors are grateful for fruitful discussion with G Birkenmeier, E Wolfrum, M Willensdorfer and H Zohm. GF Harrer is a fellow of the Friedrich Schiedel Foundation for Energy Technology. This work has been carried out within the framework of the EUROfusion Consortium and has received funding from the Euratom research and training programme 2014-2018 and 2019-2020 under grant agreement No 633053. The views and opinions expressed herein do not necessarily reflect those of the European Commission.

\section*{References}

\bibliography{cit}

\end{document}